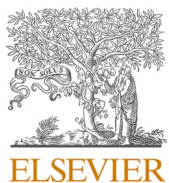
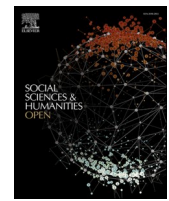

# The coronavirus trade-off – Life vs. economy: Handling the trade-off rationally and optimally

Ali Zeytoon-Nejad [a,*], Tanzid Hasnain [b]

[a] *Wake Forest University, USA*
[b] *North Carolina State University, USA*

ABSTRACT:

The recent coronavirus outbreak has made governments face an inconvenient trade-off choice, i.e. the choice between saving lives and saving the economy, forcing them to make immensely consequential decisions among alternative courses of actions without knowing what the ultimate results would be for the society as a whole. This paper attempts to frame the coronavirus trade-off problem as an economic optimization problem and proposes mathematical optimization methods to make rationally optimal decisions when faced with trade-off situations such as those involved in managing through the recent coronavirus pandemic. The framework introduced and the method proposed in this paper are on the basis of the theory of rational choice at a societal level, which assumes that the government is a rational, benevolent agent that systematically and purposefully takes into account the social marginal costs and social marginal benefits of its actions to its citizens and makes decisions that maximize the society's well-being as a whole. We approach solving this trade-off problem from a static as well as a dynamic point of view. Finally, we provide several numerical examples clarifying how the proposed framework and methods can be applied in the real-world context.

## 1. Introduction

Coronavirus disease 2019 (aka COVID-19) is an infectious disease caused by severe acute respiratory syndrome coronavirus 2 (SARS-CoV-2) (WHO, 2020). Although the origin of the virus is not still known as of the August of 2021, the disease was first identified in December 2019 in Wuhan, China (Hui D. et al., 2020). Since then, the virus has rapidly spread globally, leading to the coronavirus pandemic, in such a drastic way that as of August 5th, 2021, more than 200 million cases have been reported in more than 200 countries, and more than 4,200,000 have died due to this virus within almost 20 months since the first case was identified (Johns Hopkins University Center for Systems Science and Engineering, 2021).

The virus is spread among people very easily when they are in close proximity and its spread primarily occurs through small droplets produced during coughing, sneezing, talking, or even breathing (Stadnytskyi et al., 2020). In light of this fact, preventive measures to reduce the chances of infection including staying at home, avoiding crowded places, and practicing social-distancing strategies are introduced by epidemiologists as effective ways to reduce the contact of infected persons with large crowds by closing schools, universities, and workplaces, restricting and banning travel, and cancelling large public gatherings such as sports events, which all caused the US economy to move fast into a serious recession in 2020. According to a report by Washington Post on April 09, 2020, more than 17 million Americans filed for unemployment benefits within only four weeks prior to the report's date, which showed a swift and unprecedented decline in the U. S. economy that the nation had decided was necessary to battle the deadly coronavirus by keeping as many individuals as possible at home. In a recent short note entitled "The covid-19 recession of 2020", Greg Mankiw (2020) has described three unusual features for the coronavirus recession. He states that these three features have to do with its unusual cause (an extremely infectious and dangerous virus), its exceptional speed and depth (e.g., the unemployment rate in April 2020 was 14.7 percent - the highest level since the Great Depression), and its unusual nature (as he puts it, it was a "recession by design" rather than a "recession by accident").[1]

The outbreak of the coronavirus caused countries to encounter an inevitable, unfortunate trade-off between lives and the economy. Crook (2020) states that "Whether we like it or not, the choice between lives and the economy is already dictating events." He mentions that the governments around the world are being forced to take actions without understanding the future implications, owing to the uncertain nature of COVID-19. He goes on to say that people are looking for a rule of thumb in the pursuit of clarity in making decisions, without thinking about the complicated trade-off this pandemic has brought forth. According to his opinion, people are trying to avoid considering the crucial trade-off between the number of lives saved and the economic harm of less work. Even though, at a first glance, the decision may seem easy – saving all lives - to a decision maker, eventually the trade-off cannot be avoided. And, finally he says "But, much as we might want to ignore it, the underlying trade-off between fighting the disease and crushing the economy can't be denied. Whether we admit it or not, it's already dictating events." This article is just one example work of the scholars across the globe who have addressed this trade-off in different ways (e. g., Cornwall, 2020; Deutsche Welle, 2020; The World Bank, 2020).

A number of studies have addressed various aspects of the coronavirus outbreak and trade-off. In particular, several papers have studied the health-vs-economy trade-off by developing the "so-called" Susceptible-Infected-Recovered (SIR) model towards identifying the optimal lockdown policies. For example, Alvarez et al. (2020), Atkeson (2020), Aum et al. (2020), Bethune & Korinek (2020), Chen et al. (2020) Jones et al. (2020) have studied optimal lockdown policies for homogeneous population. Acemoglu et al., 2020; Baqaee et al., 2020; Rampini, 2020 have approached similar policies for heterogenous population, such as age-based groups. Fajgelbaum et al. (2020) have developed a spatial SIR model to study the optimal lockdown policies for commuting networks, such as Seoul and New York City. Additionally, several surveys have been performed to characterize different effects of COVID-19. Balla et al. (2020) have found that post-lockdown delays in business re-opening can be attributed to low levels of expected demand. Lin & Meissner (2020) have performed data driven analysis to understand the health-vs-economy trade-off. Chang & Velasco (2020) have developed a two-period model to study the health-vs-economy trade-off. They have developed necessary equilibrium conditions for inter-period decision taken by the households when faced to the wage (economy) vs. suffering (health) trade-off.

To the best of our knowledge, no one has thus far studied the coronavirus trade-off as a trade-off between jobs vs. lives. The present paper aims to explicitly consider the trade-off between the number of lives saved and the number of jobs saved in a static (one-period) and dynamic (two-period) setting. The contribution of our paper is two-fold. First, we explicitly consider the intertemporal relationships between the number of jobs saved and number of lives saved as constraints to the dynamic optimization model. Second, we develop closed-form solutions for the optimal number of lives and jobs saved. Accordingly, the main objective of the present paper is to frame the coronavirus trade-off problem as an economic optimization problem and propose mathematical optimization methods that enable a rational, benevolent government to make optimal decisions with regards to trade-off situations such as those involved in the optimal handling of the recent coronavirus pandemic.

The remainder of the paper is organized as follows. In Section 2, the main discussion is provided, in which the foundations of our model are explained, and the coronavirus trade-off problem is framed as an economic optimization problem. Section 3 presents the static version of the problem at hand and provides the parametric solution to the static version of the problem using mathematical methods of constrained optimization. Section 4 presents a dynamic version of the problem under study and supplies the parametric solution to the dynamic version of the problem using dynamic mathematical methods of constrained optimization. Section 5 offers conclusions. The paper ends with multiple appendices, which provide numerical examples for the framework developed and the methods used in this paper, clarifying how the proposed framework and methods can be applied in the real-world context.

## 2. Main discussion

The framework introduced and the methods proposed in this paper are based on the theory of rational choice at a public level, which assumes that the government is a rational, benevolent economic actor that uses rationally available information, and systematically and purposefully takes into account the social marginal costs and social marginal benefits of its actions to its citizens and makes decisions that maximize the society's well-being as a whole and/or consistently minimize their losses. We place the foundation of our analysis on the basis of utilitarianism and the theory of value, as advocated by utilitarian philosophers such as Jeremy Bentham and John Stuart Mill, which take into account practical consequences of actions when making an optimal decision, and thereby, we develop a mathematical objective function and employ the economic model of Possibilities Frontier to make a constraint that represents the coronavirus trade-off relationship (by showing all the combinations of lives and jobs that can possibly be saved) for our optimization problem. Then, by considering the marginal cost and marginal benefit of each action, one can simply employ mathematical optimization methods to identify the combinations of lives and jobs that makes the society best off. We attend to solving this trade-off problem from both a static as well as a dynamic standpoint.

The coronavirus trade-off has been termed by different scholars under different titles, examples of which include Medical Penalty vs. Economic Penalty, or Medical Penalty vs. Income Penalty, or Health vs. Economy, or Lives Saved vs. Income Generated, or Deaths vs. Jobs. To simplify matters, here we call it the trade-off relationship between **Saved Lives ($Y_L$)** vs. **Saved Jobs ($Y_J$)**. A common way to model trade-off relationships in economics is to use the economic model of Possibilities Frontier (henceforth, PF). Accordingly, the number of lives saved can be placed on the horizontal axis and the number of jobs saved can be put on the vertical axis of a diagram, and a PF curve can be drawn for all the combinations of $Y_L$ and $Y_J$ which can be saved if the government is doing its best and is performing efficiently. A downward-sloping PF curve represents the trade-off relationship between the two variables. We usually use the common shape of the PF curve, which is concave with respect to the origin of the coordinate system, to represent the property of increasing opportunity costs with increased quantity of either of the two outcomes, which is more realistic (than just a straight-line PF that represent constant opportunity costs) to model real-world economic phenomena. The above-described, static (i.e. one-period) trade-off relationship between **Saved Lives** ($Y_L$) vs. **Saved Jobs** ($Y_J$) can easily be modeled by such a PF curve described above. Fig. 1 shows such a PF curve for a hypothetical economy.

In this diagram, the horizontal axis shows the number of lives saved, and the vertical axis shows the number of jobs saved. This trade-off is indeed the main reason why the decision as to "when to re-open the economy" has become a hard decision for governments across the globe to make, because the health advisors (whose objective function is primarily to maximize the number of saved lives, put simply) are asking to extend the lockdown period to "save lives", while the economic advisors (whose objective function is primarily to maximize the number of "saved jobs", put simply) are asking to revoke the lockdown and re-open the economy as soon as possible. Idealistically, we would like to save all the lives and all the jobs; however, realistically, we cannot do so due to the unfortunate existing trade-off between the two outcomes. On the other hand, it is not reasonable to try to save only one or the other, primarily due to the extremely increasing opportunity costs of each outcome as we approach either of the two end points of the PF curve.[2] Despite this, we can still find an optimal combination/allocation of the two that is best for the economy from a societal point of view. The PF curve would serve as the constraint of our trade-off optimization problem. In order to be able to find an optimal point on this PF curve that shows the optimal choice of the combination of the two variables, we need to first put a price on each human life. Then, this model can be used to answer the question of "by how many percent should the economy be partially open to maximize the overall monetary benefit to the society?"

---
[1] To elaborate further on this last feature, Greg Mankiw (2020) goes on to say that "The third unusual feature of the 2020 downturn was that it was, in a sense, intentional. The typical recession is best viewed as an accident. Some surprise event shifts aggregate supply or aggregate demand, reducing production and employment. Policymakers are eager to return the economy to normal levels of production and employment as quickly as possible. By contrast, the downturn in 2020 was a recession by design. To curb the Covid-19 pandemic, policymakers compelled changes in behavior that reduced production and employment. The pandemic itself, of course, was neither intended nor desired. But given the circumstances, a large economic downturn was arguably the best outcome that could be achieved."

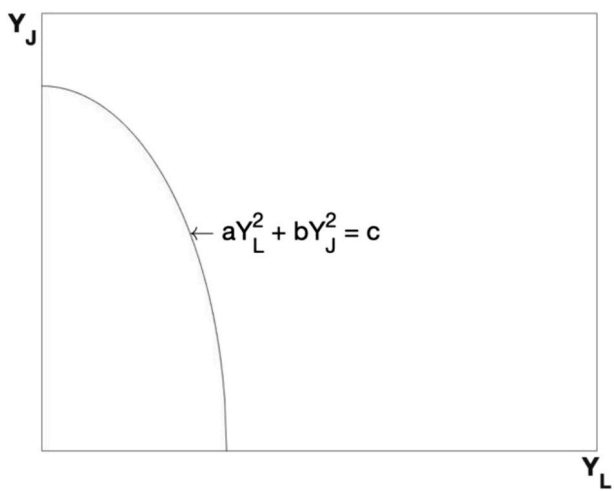

**Fig. 1.** An example possibility frontier for lives saved and jobs saved.

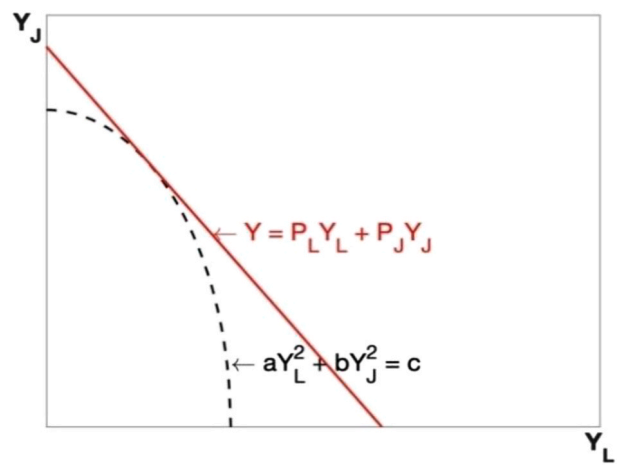

**Fig. 2.** An example possibility frontier (i.e. the constraint) for lives saved and jobs saved and a monetarized benefit curve (i.e. the objective function) for the coronavirus trade-off optimization problem.

In theory and from an ethical/deontological point of view, a human life is priceless. However, in reality and from a practical/utilitarian point of view, a human life can and should be priced, and such a valuation has become routine in Americans' daily lives nowadays, for example. Indeed, designing public policy in such a national crisis necessitates making decisions based primarily on utilitarianism and not only deontology. When legislators and regulators enact new rules and when government officials propose policies, they all need to weigh the costs and benefits associated with their decisions. One of the main questions that arises around the coronavirus trade-off is how to place a monetary value on a human life in economic terms. Only once we have done so, we can do a cost-benefit analysis to answer how many jobs we should be willing to sacrifice to save a person from dying. There are different methodologies to estimate and different estimates made for the value of a human life. For example, we know that the US government paid nearly $500,000 monetary death benefit to families when a soldier is killed in Iraq or Afghanistan (The Guardian, 2003). Additionally, behavioral economists, actuarial economists, as well as environmental economists have estimated other values for a human life, which mostly are far larger than the amount paid by the US government for the soldiers killed in war zones.[3] Regardless of the valuation methodology used to put a price on a human life, in this paper, the average value of a human life is denoted by $P_L$. We also assume that the value of a saved job is equal to its annual compensation, and it is denoted by $P_J$. Now after putting a price on a human life, we can proceed to set up an objective function for the coronavirus trade-off. If Y denotes the total monetarized benefit of the trade-off choice made for the society/economy as a whole, then the objective function of the coronavirus trade-off problem will be to maximize the total monetarized benefit of the society as a whole, which can be written as $Y = P_L \cdot Y_L + P_J \cdot Y_J$. Fig. 2 depicts a typical curve of such an objective function for a hypothetical economy along with a PF curve as the constraint of the problem.

In this diagram, the straight line on the diagram depicts the objective function of the coronavirus trade-off optimization problem. The objective function of this problem is in fact a plane with an infinite number of level curves on it representing different levels of social benefits on the schedule of our social benefit surface. Fig. 2 depicts only the highest level curve possible to be achieved that represents the maximum social benefit, which is the one that is tangent to the PF according to the principles of optimization theory. Here, we consider solely the one-period (annual) benefit of a job saved, assuming that if the job is lost this year in the economy, it can be recovered back in the economy in the next period/year (but inherent in our estimate of the value of life is the long-run cost of losing a human life).[4] Hence, one consideration that we have taken into account when framing this problem is that a job loss is a temporary loss while a life loss is a permanent loss. To see some examples of how to handle this model, you can see the numerical examples provided in the appendices of this paper.

---

[2] The concavity of the PF is a reflection of the fact that when the society strives to save all the lives, too many jobs have to be lost as too many businesses that deal with large crowds of people have to be shut down, and when the society tries to keep all the businesses running to save all the jobs, too many lives will be lost due to an extreme level of spreading of the virus. This reality can be modeled mathematically through different functional specifications, a great example of which is a quadratic functional form, which we have chosen to work with in this paper. The multipliers "a" and "b" are the parameters that determine the degree and shape of the concavity.

[3] The value of a human life has been estimated to be $9.1 million by Environmental Protection Agency in 2010, and $7.9 million by Food and Drug Administration in 2010, and $9.2 million by the Department of Transportation, 2014, and $9.6 million by the Department of Transportation in 2016. To see more information on these, you can see the following: The Globalist (2012) and TIME (2020).

By following this information and framework, we can answer many important questions in a rational manner. Examples of such crucial questions include: What are the optimal levels of $Y_L$ and $Y_J$ in the economy? How much is the maximized total monetarized benefit that this rational decision by the government can bring about for the society/economy as a whole? In Section 3, we will answer these questions for the static version of this trade-off optimization problem given its parametric constraint and objective function.[5]

## 3. The static version of the coronavirus trade-off problem

In this section, we develop the optimization model for the static version of the coronavirus trade-off problem. We produce closed-form solutions for the optimal number of lives and jobs saved and develop the equation to estimate the maximized total monetarized benefit in the presence of lives-vs-jobs trade-off. Following the discussion in section 2, we develop the optimization model for the static framework, denoted by *Stat*, as follows:

Model    *Stat*
$\underset{Y_L, Y_J}{\text{Maximize}} \quad Z_s = P_L Y_L + P_J Y_J$
s.t.    $a Y_L^2 + b Y_J^2 = c$

The Lagrangian function for the model *Stat* can be written as:

$$\mathscr{L}_s(Y_L, Y_J, \lambda) = P_L Y_L + P_J Y_J + \lambda\left(c - a Y_L^2 - b Y_J^2\right)$$

where subscript *s* of $\mathscr{L}$ denotes the static version of the model (i.e. the *Stat* model). The First Order Conditions (**FOCs**) of the above optimization problem can be stated as follows:

$$\frac{\partial \mathscr{L}_s}{\partial Y_L} = P_L - 2a\lambda Y_L = 0$$

$$\frac{\partial \mathscr{L}_s}{\partial Y_J} = P_J - 2b\lambda Y_J = 0$$

$$\frac{\partial \mathscr{L}_s}{\partial \lambda} = c - a Y_L^2 - b Y_J^2 = 0$$

By solving the system of the FOCs above, we can derive the **Optimality Condition** as follows:

$$\frac{Y_L}{Y_J} = \frac{bP_L}{aP_J}$$

The **optimal $Y_L$ and $Y_J$** derived from the FOCs can be stated as:

$$Y_L^* = \sqrt{\left(\frac{b^2 c P_L^2}{ab^2 P_L^2 + a^2 b P_J^2}\right)}$$

$$Y_J^* = \sqrt{\left(\frac{a^2 c P_J^2}{ab^2 P_L^2 + a^2 b P_J^2}\right)}$$

Thus, the *maximized total monetarized benefit* that this rational decision to be made by the government can bring about for the society/economy as a whole can be stated as:

$$Z_s^* = P_L \cdot \sqrt{\left(\frac{b^2 c P_L^2}{ab^2 P_L^2 + a^2 b P_J^2}\right)} + P_J \sqrt{\left(\frac{a^2 c P_J^2}{ab^2 P_L^2 + a^2 b P_J^2}\right)}$$

## 4. The dynamic version of the coronavirus trade-off problem

An alternative and more comprehensive way to investigate the coronavirus trade-off is to look at it from a dynamic viewpoint and take into account the intertemporal interrelations between the variables involved in this trade-off. Continuing our context from the previous section, this section incorporates the dynamic aspects of this trade-off relationship into the optimization problem. Now, assume that we are considering the same problem over a two-period time horizon (i.e. two years). Additionally, suppose that it has been understood that there are two important dynamic aspects to the decision at hand, which are as follows:

a) There is a dynamic trade-off between the number of jobs saved this year ($Y_{J1}$) and the number of lives saved next year ($Y_{L2}$), meaning that when the former goes up, the latter goes down, primarily due to the fact that, over time, even more jobs will be created in a network-like, exponential manner, and the extent of social contact, exposure, mobility, and therefore, the spread of the virus increases over time, resulting in a lower number of lives saved in the next year. Additionally, assume that it has been estimated that the rate of this dynamic trade-off over time follows the following equation (where the numerical subscripts denote the respective time period numbers, except for a, b, and c, whose numerical subscripts indicate the respective constraint numbers): $a_1 Y_{L2}^2 + b_1 Y_{J1}^2 = c_1$

b) Furthermore, there is a dynamic trade-off between the number of lives saved this year ($Y_{L1}$) and the number of jobs saved next year ($Y_{J2}$), meaning that when the former goes up, the latter goes down. This is because requiring people to stay at home this year causes their income to go down, which in turn decreases their demand for goods and services next year, which in turn causes fewer jobs needed to create fewer goods and services next year. Additionally, assume that it has been estimated that the rate of this dynamic trade-off follows the following equation: $a_2 Y_{L1}^2 + b_2 Y_{J2}^2 = c_2$

Notice that this optimization problem is now a 'dynamic' decision, so it has to deal with the time value of money. Assume that the social discounting rate for this problem is denoted by *i* (note that *i* is a decimal number such as 0.02) for the sake of our discounting computations, and that, for simplicity, the social discounting rate does not influence any of the constraints mentioned above, and it just influences our objective function. Now, one can use the above context to answer the same questions that we answered in the previous section, but now we can answer them after taking into account the temporal effects existing in the context of the optimization problem.

---

[4] Here, we are trying to model an economic phenomenon, so we need to simplify matters by making several simplifying assumptions, since our purpose here is to put together a workable, comprehendible mathematical-economic problem. In the real world, modeling such a trade-off relationship would involve many other subtleties and complexities such as its dynamic aspects, which all need to be taken into account when making a decision about the optimal degree of the partial operation of the economy and when finding the best combination of saved jobs and saved lives, whose discussion is beyond the scope of this section of the paper, and we attend to that dimension of the problem in Section 4.

[5] We solve the problems under study in this paper first parametrically, so that we can preserve the positivity of our approach and the objectivity of our model, method, and analysis. We leave the choice of the parameters $P_L$ and $P_J$ (each of which can be the subject matter of a standalone research study *per se*) to the user of our model, since the choice of these parameters is the normative and subjective part of the discussion, which we as economists prefer to avoid dealing with at this stage of our modeling. Accordingly, once the model and problem is solved parametrically, it will have a solution for any set of subjective $P_L$ and $P_J$ that the user would tend to put on a human life and a job, regardless of their philosophies of virtues and vices.

In this section, we develop the optimization model for the dynamic version of the coronavirus trade-off problem. We produce closed-form solutions for the optimal number of lives and jobs saved in each period and develop the equation to estimate the maximized total monetarized benefit in the presence of lives-vs-jobs trade-off. The dynamic optimization model, which is called *Dynam*, takes the following form:

Model    Dynam

$$\underset{Y_{L1}, Y_{L2}, Y_{J1}, Y_{J2}}{\text{Maximize}} \quad Z_d = P_{L1} Y_{L1} + P_{J1} Y_{J1} + \frac{P_{L2} Y_{L2}}{1+i} + \frac{P_{J2} Y_{J2}}{1+i}$$

$$\text{s.t.} \quad a_1 Y_{L2}^2 + b_1 Y_{J1}^2 = c_1$$

$$a_2 Y_{L1}^2 + b_2 Y_{J2}^2 = c_2$$

The Lagrangian function for the model *Dynam* can be written as:

$$\mathcal{L}_d(Y_{L1}, Y_{L2}, Y_{J1}, Y_{J2}, \lambda_1, \lambda_2) = P_{L1} Y_{L1} + P_{J1} Y_{J1} + \frac{P_{L2} Y_{L2}}{1+i} + \frac{P_{J2} Y_{J2}}{1+i}$$
$$+ \lambda_1 \left( c_1 - a_1 Y_{L2}^2 - b_1 Y_{J1}^2 \right) + \lambda_2 \left( c_2 - a_2 Y_{L1}^2 - b_2 Y_{J2}^2 \right)$$

where subscript $d$ of $\mathcal{L}$ indicates that the model is a dynamic one. The First Order Conditions (**FOCs**) can be stated as follows:

$$\frac{\partial \mathcal{L}_d}{\partial Y_{L1}} = P_{L1} - 2a_2 \lambda_2 Y_{L1} = 0$$

$$\frac{\partial \mathcal{L}_d}{\partial Y_{L2}} = \frac{P_{L2}}{1+i} - 2a_1 \lambda_1 Y_{L2} = 0$$

$$\frac{\partial \mathcal{L}_d}{\partial Y_{J1}} = P_{J1} - 2b_1 \lambda_1 Y_{J1} = 0$$

$$\frac{\partial \mathcal{L}_d}{\partial Y_{J2}} = \frac{P_{J2}}{1+i} - 2b_2 \lambda_2 Y_{J2} = 0$$

$$\frac{\partial \mathcal{L}_d}{\partial \lambda_1} = c_1 - a_1 Y_{L2}^2 - b_1 Y_{J1}^2 = 0$$

$$\frac{\partial \mathcal{L}_d}{\partial \lambda_2} = c_2 - a_2 Y_{L1}^2 - b_2 Y_{J2}^2 = 0$$

By solving the system of the FOCs above, the following **Optimality Conditions** are solved for:

$$\frac{Y_{L1}}{Y_{J2}} = \frac{b_2(1+i)P_{L1}}{a_2 P_{J2}}$$

$$\frac{Y_{L2}}{Y_{J1}} = \frac{b_1 P_{L2}}{a_1(1+i)P_{J1}}$$

The **optimal** $Y_{L1}, Y_{L2}, Y_{J1}, Y_{J2}$ derived from the FOCs can be expressed as:

$$Y_{L1}^* = \sqrt{\left( \frac{c_2 b_2 (1+i)^2 P_{L1}^2}{a_2 b_2^2 (1+i)^2 P_{L1}^2 + b_2 a_2^2 P_{J2}^2} \right)}$$

$$Y_{L2}^* = \sqrt{\left( \frac{c_1 b_1^2 P_{L2}^2}{a_1 b_1^2 P_{L2}^2 + b_1 a_1^2 (1+i)^2 P_{J1}^2} \right)}$$

$$Y_{J1}^* = \sqrt{\left( \frac{c_1 a_1^2 (1+i)^2 P_{J1}^2}{a_1 b_1^2 P_{L2}^2 + b_1 a_1^2 (1+i)^2 P_{J1}^2} \right)}$$

$$Y_{J2}^* = \sqrt{\left( \frac{c_2 a_2^2 P_{J2}^2}{a_2 b_2^2 (1+i)^2 P_{L1}^2 + b_2 a_2^2 P_{J2}^2} \right)}$$

Thus, the *maximized total monetarized benefit* that this rational decision to be made by the government can bring about for the society/economy as a whole can be shown as:

$$Z_d^* = P_{L1} \sqrt{\left( \frac{c_2 b_2 (1+i)^2 P_{L1}^2}{a_2 b_2^2 (1+i)^2 P_{L1}^2 + b_2 a_2^2 P_{J2}^2} \right)}$$
$$+ P_{J1} \sqrt{\left( \frac{c_1 a_1^2 (1+i)^2 P_{J1}^2}{a_1 b_1^2 P_{L2}^2 + b_1 a_1^2 (1+i)^2 P_{J1}^2} \right)}$$
$$+ \frac{P_{L2}}{1+i} \sqrt{\left( \frac{c_1 b_1^2 P_{L2}^2}{a_1 b_1^2 P_{L2}^2 + b_1 a_1^2 (1+i)^2 P_{J1}^2} \right)}$$
$$+ \frac{P_{J2}}{1+i} \sqrt{\left( \frac{c_2 a_2^2 P_{J2}^2}{a_2 b_2^2 (1+i)^2 P_{L1}^2 + b_2 a_2^2 P_{J2}^2} \right)}$$

## 5. Remarks and conclusion

In this section, a few important remarks are made, and afterwards, a conclusion will be drawn. Here is a list of considerations that should be taken into account with regard to the coronavirus trade-off:

- This paper is not to demean or degrade the deontological position that one can take with respect to the coronavirus trade-off in general and to the value of life in particular. In fact, even extreme polar deontological positions (such as putting an infinite value on a human life) can be accommodated as a special case by the economic model that we have proposed. Rather, the present paper suggests that the best that can be done with regard to managing this social trade-off optimally is to look at the issue from a utilitarian point of view and strive to solve it practically through the decisions and actions that bring about the least social loss to be incurred by the society, instead of leaving the problem unsolved and solely having discussions about the value of a human life being infinite. Ultimately, the main mission of the government as an economic institution facing a pandemic crisis is to make decisions and take actions at a public scale by designing and conducting public policies. To that end, the best a rational government can do is to act rationally and make rationally optimal decisions, instead of solely opening up a social, philosophical debate talking only about the deontological aspects of the matters, while making decisions baselessly and taking actions cluelessly on what combination of lives and jobs should be saved in practice.

- It is important to admit that there are many aspects to the political economy of handling a pandemic trade-off such as the coronavirus trade-off. Ultimately, governments may be hesitant to explicitly articulate the utilitarian basis of their decisions made and actions taken, if they know that they will be criticized by the public for the price that they have put on a human life in their decision-making process, for example, for doing a cost-benefit analysis over human lives. In such cases, governments usually opt to not be transparent about the bases of their decision-making, which is understandable to some extent when one takes into account that politicians maximize their own objective functions (e.g., maximizing the number of their votes to be received in future elections), but it should not cause them to make baseless decisions that do not maximize the society's well-being. However, instead, governments should communicate well to the public the bases of their decisions and actions, and make it clear that what is being done is the best that can be done given the existing emergency situation, and that this is certainly much better than not having any model or plan in place for handling the trade-off optimally, which definitely causes more losses to the society.

- To identify and estimate the equation of the constraint of our model, there should be a decent group of labor economists (especially, macro-level labor economists), econometricians, biologists, epidemiologists, and other related experts who can work together and in collaboration to estimate the trade-off equations (i.e. the constraints of the problem) correctly.

- Effective preventative practices such as wearing masks and maintaining social distancing will shift the curve outside (i.e. upward and to the right), enabling the economy to get more of both outcomes, that is, more saved lives and more saved jobs, but the trade-off still remains although at a higher level of outcomes. However, the invention and administration of an effective vaccine or the invention of an effective treatment to the disease would eliminate the trade-off entirely.
- As mentioned earlier, technological advancements in preventative practices can cause the PF schedule shift outward. If the effect of the technological advancements is proportionate to both outcomes (i.e., saved lives and saved jobs), the PF will shift outward parallelly. If the effect of technological advancements is disproportionate with respect to the two outcomes, the PF will shift outward unparallelly.
- As the level of preventative practices in society may change over time (they may weaken or strengthen over time), and as the level of vaccination evolve over time (it would weaken the trade-off relationship), which in turn can shift the PF inward or outward, the optimal decision about the trade-off must be re-considered and adjusted once in a while (especially, after any significant shift in the PF) accordingly to the new schedule of the PF, in order to ensure that the society is at its new optimal point with respect to the trade-off choice at any point in time and after any significant shifts or changes in the PF schedule.
- Certainly, there will be a geographical heterogeneity in the shape of the PF curve, and different locations may face different PF constraints, depending on locally heterogeneous characteristics and factors. We assume that the PF constraint discussed in this paper was a constraint at an aggregate level for the economy as a whole.
- There are many different factors playing a role in this context, such as medical factors, political-economic factors, and behavioral factors. For example, there are many major, important political-economic aspects to the proper handling of the COVID-19 recession, which limit the extent to which the government can handle the crisis optimally, but we find it beyond the scope of the present paper to discuss the political economy of the coronavirus pandemic and trade-off, so we have not included those important aspects of the analysis in our exposition of the problem.

In this paper, we have developed a static (single-period) and a dynamic (two-period) optimization framework to optimally handle the lives-vs-jobs trade-off generated as a result of COVID-19. In the static framework, we have considered a hypothetical Possibility Frontier curve which constrains an economy to maximize the number of lives saved and the number of jobs saved simultaneously. On the other hand, we have considered cross-temporal constraints between the number of lives saved and the number of jobs saved in the dynamic framework. For both static and dynamic frameworks, we have developed the functional formulations for the optimal number of lives and jobs saved for a given set of parameters (including cost of a life, cost of a job, and constraint parameters), as well as the maximized total monetarized benefit. The results for numerical examples of this analysis can be found in Appendices 1, 2, and 3.

An obvious direction for future work is the parameterization. In this paper, we have focused on developing the frameworks for an economy in the presence of lives-vs-jobs trade-off in a world affected by COVID-19. While we have provided several reasonings for cost-parameter selections (including cost of a life and cost of a job), much work can be done on the parameterization and estimation of these cost parameters and constraint parameters, both static ones (that is, $a, b,$ and $c$) and dynamic ones (i.e. $a_1, b_1, c_1, a_2, b_2,$ and $c_2$). Moreover, the two-period dynamic framework can be extended to a T-period framework. Additionally, and very interestingly, this framework can be used as a basis to empirically estimate how much price/value different governments have implicitly placed on jobs and lives in their decision-making, policy-formulation, and actions through looking at the data of the lives lost and the jobs lost in their respective jurisdictions and territories.

**CRediT authorship contribution statement**

**Ali Zeytoon-Nejad:** Conceptualization, Methodology, framing optimization problems, solving the mathematical problems, Visualization, Writing – original draft, Writing – review & editing. **Tanzid Hasnain:** Conceptualization, Methodology, framing optimization problems, solving the mathematical problems, Visualization, Writing – original draft, Writing – review & editing, Both authors contributed equally.

**Appendices.**

*Appendix 1: A Simple Example Representation of the Coronavirus Trade-off Problem with 6 Points on the PF Curve*

Fig. 3 depicts the PF curve for a hypothetical economy. Six points have been selected with the following coordinates of $Y_L$ and $Y_J$, respectively, where $(Y_L = 0, Y_J = 10)$ for Point (1), **(0.2, 9.8)** for Point (2), **(0.4, 9.2)** for Point (3), **(0.6, 8)** for Point (4), **(0.8, 6)** for Point (5), and **(1, 0)** for Point (6).

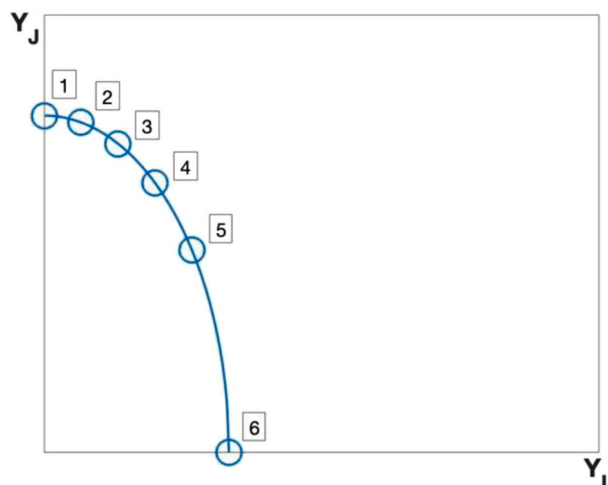

**Fig. 3.** An example representation of the PF curve.

The horizontal axis is calibrated in terms of millions of lives saved, and the vertical axis is calibrated in the terms of millions of jobs saved. In this question, assume that the average value of a human life is $1,000,000, and that the value of a saved job is $60,000 (and recall that we only look at one-period benefit of a job saved, arguing that a job loss is a temporary loss while a life loss is a permanent loss). Suppose that $P_L$ denotes the value of a human life, $P_J$ denotes the value of a saved job, $Y_L$ denotes the number of lives saved, $Y_J$ denotes the number of jobs saved, and Z denotes the total monetarized benefit of the trade-off choice made for the society/economy as a whole, where $Z = P_L \cdot Y_L + P_J \cdot Y_J$. The *total monetarized benefit* for the six points mentioned above are as follows:

**Point 1:** $Z = \$600$ billion
**Point 2:** $Z = \$788$ billion
**Point 3:** $Z = \$952$ billion
**Point 4:** $Z = \$1,080$ billion
**Point 5:** $Z = \$1,160$ billion
**Point 6:** $Z = \$1,000$ billion

Thus, Point (5), where the number of lives saved is 6 million and the number of jobs saved is 800,000, is the optimal point for the economy to operate on and it produces the highest total monetarized benefit of $1.16 trillion for the economy.

***Appendix 2:*** *A Numerical Example for the Static Version of the Coronavirus Trade-off Problem with an Equation for the PF Curve*

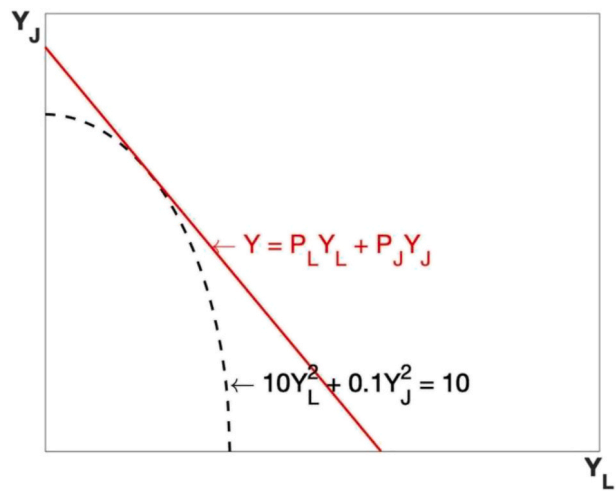

**Fig. 4.** A numerical example of the possibility frontier described in Section 3.

Let us consider a numerical instance of the Possibility Frontier as depicted in Fig. 4. The optimization model for this static framework can be formulated as the following:

*Model     Stat*
*Maximize*   $Z_s = P_L Y_L + P_J Y_J$
*s.t.*         $10\, Y_L^2 + 0.1 Y_J^2 = 10$

where $P_L = \$1,000,000$ and $P_J = \$60,000$. The Lagrangian function for the model *Stat* can be written as:

$$\mathscr{L}_s(Y_L, Y_J, \lambda) = P_L Y_L + P_J Y_J + \lambda\left(10 - 10\, Y_L^2 - 0.1 Y_J^2\right)$$

where the subscript *s* of $\mathscr{L}$ indicates that this model is a static one. The First Order Conditions (FOCs) of the above optimization problem are as follows:

$$\frac{\partial \mathscr{L}_s}{\partial Y_L} = P_L - 20\lambda Y_L = 0$$

$$\frac{\partial \mathscr{L}_s}{\partial Y_J} = P_J - 0.2\lambda Y_J = 0$$

$$\frac{\partial \mathscr{L}_s}{\partial \lambda} = c - 10 Y_L^2 - 0.1 Y_J^2 = 0$$

By solving the system of the FOCs above, the optimality condition can be derived as follows:

$$\frac{Y_J}{Y_L} = \frac{20 P_J}{0.2 P_L} = 100 \frac{P_J}{P_L}$$

For $P_L = \$1,000,000$, $P_J = \$60,000$, the optimality condition reduces to $\frac{Y_J}{Y_L} = 6$. Thus, to maximize the total monetarized benefit of the trade-off choice made for the society as a whole, the total number of jobs saved should be six times the total number of lives saved.

Accordingly, the optimal values of $Y_L$ and $Y_J$ obtained from the FOCs are as follows:

$Y_L^* = 0.8575$ million lives to be saved
$Y_J^* = 5.1450$ million jobs to be saved

Therefore, the maximized total monetarized benefit that this rational decision to be made by the government can bring about for the society/economy as a whole is roughly equal to:

$Z_s^* = \$1.166$ trillion

As another example, for $P_L = \$500,000$, $P_J = \$120,000$, the optimal values of $Y_L$ and $Y_J$ and the total monetarized benefit will be roughly equal to:

$Y_L^* = 0.3846$ million lives to be saved
$Y_J^* = 9.2308$ million jobs to be saved
$Z_s^* = \$1.3$ trillions

**Appendix 3:** *A Numerical Example for the Dynamic Version of the Coronavirus Trade-off Problem*

Let us consider two numerical instances for the two constraints of the *Dynam* model as follows:

$0.2\, Y_{L2}^2 + Y_{J1}^2 = 1$

$Y_{L1}^2 + 0.1 Y_{J2}^2 = 2$

The dynamic optimization model can then be formulated as the following:

*Model    Dynam*

Maximize $\quad Z_d = P_{L1} Y_{L1} + P_{J1} Y_{J1} + \dfrac{P_{L2} Y_{L2}}{1+i} + \dfrac{P_{J2} Y_{J2}}{1+i}$

s.t. $\quad\quad 0.2\, Y_{L2}^2 + Y_{J1}^2 = 1$

$\quad\quad\quad Y_{L1}^2 + 0.1 Y_{J2}^2 = 2$

where, $P_{L1} = P_{L2} = \$1,000,000$ and $P_{J1} = P_{J2} = \$60,000$, for simplicity. The Lagrangian function can be written as:

$$\mathscr{L}_d(Y_{L1}, Y_{L2}, Y_{J1}, Y_{J2}, \lambda_1, \lambda_2) = P_{L1} Y_{L1} + P_{J1} Y_{J1} + \frac{P_{L2} Y_{L2}}{1+i} + \frac{P_{J2} Y_{J2}}{1+i} + \lambda_1 \left(1 - 0.2\, Y_{L2}^2 - Y_{J1}^2\right) + \lambda_2 \left(2 - Y_{L1}^2 - 0.1 Y_{J2}^2\right)$$

where the subscript $d$ of $\mathscr{L}$ indicates that the model is a dynamic one. The First Order Conditions (FOCs) can be derived as follows:

$\dfrac{\partial \mathscr{L}_d}{\partial Y_{L1}} = P_{L1} - 2\lambda_2 Y_{L1} = 0$

$\dfrac{\partial \mathscr{L}_d}{\partial Y_{L2}} = \dfrac{P_{L2}}{1+i} - 0.4 \lambda_1 Y_{L2} = 0$

$\dfrac{\partial \mathscr{L}_d}{\partial Y_{J1}} = P_{J1} - 2\lambda_1 Y_{J1} = 0$

$\dfrac{\partial \mathscr{L}_d}{\partial Y_{J2}} = \dfrac{P_{J2}}{1+i} - 0.2 \lambda_2 Y_{J2} = 0$

$\dfrac{\partial \mathscr{L}_d}{\partial \lambda_1} = 1 - 0.2\, Y_{L2}^2 - Y_{J1}^2 = 0$

$\dfrac{\partial \mathscr{L}_d}{\partial \lambda_2} = 2 - Y_{L1}^2 - 0.1 Y_{J2}^2 = 0$

By solving the system of the FOCs above, the following optimality conditions are obtained:

$\dfrac{Y_{L1}}{Y_{J2}} = \dfrac{0.1(1+i) P_{L1}}{P_{J2}}$

$\dfrac{Y_{L2}}{Y_{J1}} = \dfrac{P_{L2}}{0.2(1+i) P_{J1}}$

For $P_{L1} = P_{L2} = \$1,000,000$ and $P_{J1} = P_{J2} = \$60,000$, the optimality conditions reduce to:

$$\frac{Y_{L1}}{Y_{J2}} = 1.7$$

$$\frac{Y_{L2}}{Y_{J1}} = 81.7$$

Thus, to maximize the total monetarized benefit of the trade-off choice made for the economy as a whole, the total number of lives saved in Period 1 must be set to be 1.7 times the total number of jobs saved in Period 2, and the total number of lives saved in Period 2 must be set to be 81.7 times the total number of jobs saved in Period 1.

Accordingly, the optimal values of $Y_{L1}, Y_{L2}, Y_{J1}, Y_{J2}$ obtained from the FOCs are as follows:

$Y_{L1}^* = 1.3903$ millions of lives saved in Period 1
$Y_{L2}^* = 2.2352$ millions of lives saved in Period 2
$Y_{J1}^* = 0.0273$ millions of jobs saved in Period 1
$Y_{J2}^* = 0.8178$ millions of jobs saved in Period 2

Thus, the maximized total monetarized benefit that this rational decision to be made by the government can bring about for the society/economy as a whole equals:

$Z_d^* = \$3.631$ trillion

As another example, for $P_{L1} = P_{L2} = \$500,000$ and $P_{J1} = P_{J2} = \$120,000$, the optimal values of $Y_{L1}, Y_{L2}, Y_{J1}, Y_{J2}$, and the total monetarized benefit would change to:

$Y_{L1}^* = 1.1345$ millions of lives saved in Period 1
$Y_{L2}^* = 2.2227$ millions of lives saved in Period 2
$Y_{J1}^* = 0.1088$ millions of jobs saved in Period 1
$Y_{J2}^* = 2.6696$ millions of jobs saved in Period 2
$Z_d^* = \$1.984$ trillions